\documentclass[aps,preprintnumbers,eqsecnum,amsmath,amssymb,showpacs,
nofootinbib]{revtex4}
\usepackage{graphicx}
\usepackage{dcolumn}
\usepackage{bm}

\begin{document} 
\title{\boldmath Are there narrow flavor-exotic tetraquarks in
large-$N_c$ QCD?}
\author{Wolfgang Lucha$^a$, Dmitri Melikhov$^{a,b,c}$, Hagop Sazdjian$^d$}
\affiliation{
$^a$Institute for High Energy Physics, Austrian Academy of Sciences,
Nikolsdorfergasse 18, A-1050 Vienna, Austria\\ 
$^b$D.~V.~Skobeltsyn Institute of Nuclear Physics, M.~V.~Lomonosov
Moscow State University, 119991, Moscow, Russia\\
$^c$Faculty of Physics, University of Vienna, Boltzmanngasse 5,
A-1090 Vienna, Austria\\ 
$^d$Institut de Physique Nucl\'eaire, CNRS-IN2P3, Universit\'e
Paris-Sud,   Universit\'e Paris-Saclay, 91406 Orsay, France}
\date{\today}

\begin{abstract}
A salient feature shared by all tetraquark candidates observed in
experiment is the absence of flavor-exotic states of the type
$\bar a b\bar c d$, with four different quark flavors. 
This phenomenon may be understood from the properties of large-$N_c$
QCD: On the one hand, consistency conditions for flavor-exotic
Green functions, potentially containing these tetraquark poles,
require the existence of two tetraquarks $T_A$ and $T_B$: each of
them should decay dominantly via a single two-meson channel,   
$T_A\to M_{\bar a b}M_{\bar c d}$ and
$T_B\to M_{\bar a d}M_{\bar c b}$, with suppressed rates 
$T_A\to M_{\bar a d}M_{\bar c b}$ and
$T_B\to M_{\bar a b}M_{\bar c d}$. 
On the other hand, we have at hand only one diquark-antidiquark
flavor structure $(\bar a \bar c)(b d)$ that might produce a
compact tetraquark bound state. Taking into account that the
diquark-antidiquark structure is the only viable candidate
for a compact tetraquark state, one concludes that it is impossible
to have two different narrow tetraquarks decaying dominantly into
different two-meson channels. 
This contradiction suggests that large-$N_c$ QCD does not support
the existence of narrow flavor-exotic tetraquarks.
This argument does not rule out the possible existence of broad
molecular-type flavor-exotic states, or of molecular-type bound
states lying very close to the two-meson thresholds.
\end{abstract}

\pacs{11.15.Pg, 12.38.Lg, 12.39.Mk, 14.40.Rt}
\maketitle

\section{Motivation} \label{s1}

In recent years, many narrow near-threshold hadron resonances
which have a favourable interpretation as tetraquark and pentaquark  
hadrons (i.e., hadrons with minimal parton configurations consisting
of four and five quarks, respectively) have been observed
experimentally \cite{ali,Olsen:2017bmm}. An intriguing feature shared
by all exotic candidates is the absence of states with 
flavor-exotic structure, i.e., with four different quark flavors,
which cannot be realized in a single ordinary hadron. The only
flavor-exotic tetraquark candidate $X(5568)$ of D0 \cite{D0} was not
confirmed by LHCb \cite{LHCb}, CMS \cite{CMS}, CDF \cite{CDF}, and
ATLAS \cite{ATLAS}.
Lattice calculations also seem to rule out the existence of
tetraquarks with such a structure \cite{Ikeda:2013vwa}, at least
with the quark content ($cs\bar u\bar d$).
In this paper, we attempt to understand this phenomenon from the
large-$N_c$ perspective and give arguments why in large-$N_c$ QCD
no narrow compact flavor-exotic states may exist. 

QCD with a large number of colors $N_c$ (i.e., ${\rm SU}(N_c)$ gauge
theory for large $N_c$, with quarks in the fundamental representation)
with a simultaneously decreasing coupling $\alpha_s\sim1/N_c$
\cite{largeNc1,largeNc2} has proven to be a useful theoretical tool
to explain the essential properties of hadron interactions and, in
particular, the properties of possibly existing tetraquark and
pentaquark hadrons \cite{coleman,weinberg,knecht,cohen,maiani,
Esposito:2016noz,lms_prd,lms_epjc,maiani2}. 

Recently, we have formulated rigorous criteria to be satisfied by the
four-point Green functions of bilinear quark currents appropriate for
tetraquark poles \cite{lms_prd,lms_epjc}: any diagram which
contributes to the potential tetraquark pole in the Mandelstam
variable $s$, at $s=M_T^2$, where $M_T$ is the tetraquark mass,
should satisfy the following two almost self-evident criteria:
(i) The diagram should have a nontrivial (i.e., non-polynomial)
dependence on $s$. (ii) It should support four-quark
intermediate states and corresponding cuts starting at
$s=(m_1+m_2+m_3+m_4)^2$, where $m_i$ are the masses of the quarks
forming the tetraquark bound state. The presence or absence of this
cut is established by solving the Landau equations for the
corresponding diagram \cite{landau}. Hereafter, we refer to diagrams
which satisfy these criteria as {\it tetraquark-phile} diagrams.

Here, we take a closer look at the four-point Green functions of
quark bilinear currents (omitting spin and Lorentz indices, which
do not play a fundamental role) of exotic flavor content. We show
that the tetraquark-phile diagrams have a cylinder topology: For the
{direct} ({\cal D}) Green functions
$\langle T\{j_{\bar ab}j_{\bar cd}j^\dagger_{\bar ab}
j^\dagger_{\bar cd}\}\rangle$ and 
$\langle T\{j_{\bar ad}j_{\bar cb}j^\dagger_{\bar ad}
j^\dagger_{\bar cb}\}\rangle$, the $N_c$-leading contributions are
given by diagrams with multiple nonintersecting gluon exchanges lying
on the tube; all these diagrams behave like $O(N_c^0)$. Cylinder
diagrams with $k$ gluonic handles, which avoid intersection of gluon
lines on the tube, behave like $O(N_c^{-2k})$. For the {recombination}
({\cal R}) Green functions 
$\langle T\{j_{\bar ab}j_{\bar cd}j^\dagger_{\bar ad}
j^\dagger_{\bar cb}\}\rangle$, the $N_c$-leading tetraquark-phile
diagrams are those containing one gluonic handle and they behave
like $O(N_c^{-1})$; adding $k$ gluonic handles leads to suppressed
diagrams of order $O(N_c^{-1-2k})$. The above classification of
behaviors does not take into account the eventual presence of quark
loops. The insertion of a quark loop inside gluon lines lowers the
scaling power in $N_c$ by one unit.

Because of the different $N_c$ scalings (even powers for {\cal D}
and odd powers for {\cal R}) of the tetraquark-phile Green functions,
one finds that narrow tetraquarks with a fixed mass at large $N_c$
may emerge only in pairs, $T_A$ and $T_B$, each of them decaying via
one preferred two-meson channel \cite{lms_prd,lms_epjc}. Next,
since the only viable flavor structure of narrow compact tetraquark
bound states, resulting from confinement, is the diquark
structure \cite{Schafer:1993ra,Jaffe:2003sg,Nussinov:2003ex,
Shuryak:2003zi,Maiani:2004vq,Brodsky:2014xia,Karliner:2017qjm,
Eichten:2017ffp,Maiani:2017kyi} and in the flavor-exotic case one
has only one flavor diquark-antidiquark combination,
$(\bar a\bar c)(bd)$ (product of antisymmetric representations in
color space), one encounters a contradiction with the requirement of
the existence of two different tetraquarks. One thus has to conclude
that, without the presence of a dynamical fine-tuning mechanism,
narrow flavor-exotic compact states do not exist in large-$N_c$ QCD.

\section{Narrow flavor-exotic tetraquark states at large $N_c$}
\label {s2}

We consider the case of possibly existing narrow flavor-exotic
tetraquark states and study the properties of tetraquark-phile
diagrams for direct Green functions 
$\langle T\{j_{\bar ab}j_{\bar cd}j^\dagger_{\bar ab}
j^\dagger_{\bar cd}\}\rangle$ and 
$\langle T\{j_{\bar ad}j_{\bar cb}j^\dagger_{\bar ad}
j^\dagger_{\bar cb}\}\rangle$ and recombination Green functions 
$\langle T\{j_{\bar ab}j_{\bar cd}j^\dagger_{\bar ad}
j^\dagger_{\bar cb}\}\rangle$. 

\subsection{Direct 4-point Green functions}
\label{s2A}

\begin{figure}[!h]
\centering
\includegraphics[height=4.0cm]{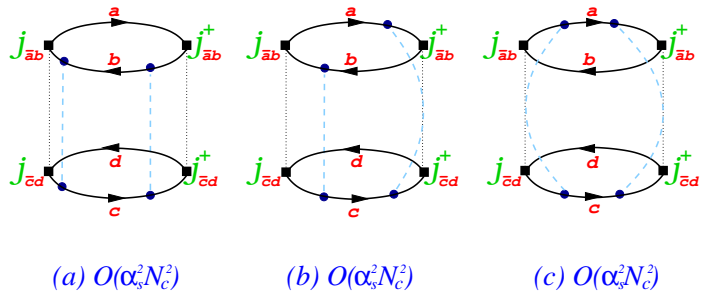}
\caption{\label{Fig:m1}
  Tetraquark-phile diagrams for the direct amplitude with two gluon
  exchanges.
  All diagrams have cylinder topology and two color loops and are of
  order $\alpha_s^2 N_c^2$.
  Light-blue dashed lines denote planar gluons (lying on the cylinder
  tube); grey dotted lines indicate the contours of the tube.}
\end{figure}

According to the formulated criteria for selecting tetraquark-phile
diagrams (which give necessary, but not sufficient conditions for the
existence of tetraquark poles), the lowest-order diagrams in
$\alpha_s$ are cylinder diagrams of Fig.~\ref{Fig:m1} with two-gluon
exchanges between the quark loops. They have cylinder topology,
contain two color loops and are of order
$O(\alpha_s^2 N_c^2)=O(N_c^0)$ at large $N_c$. 

\begin{figure}[!h]
\includegraphics[height=3.8cm]{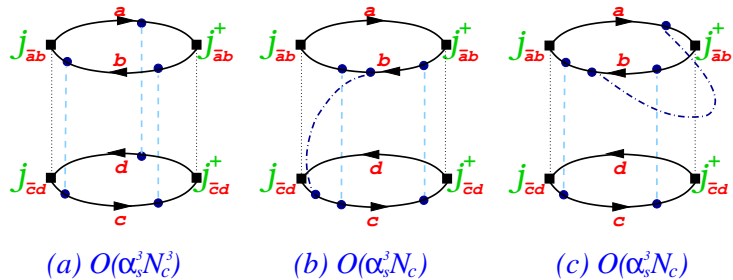}
\caption{\label{Fig:m2}
  Diagrams for a direct amplitude, appropriate for flavor-exotic
  tetraquarks, of order $O(\alpha_s^3)$, with three gluon exchanges.
  Light-blue dashed lines denote planar gluons (lying on the cylinder
  tube); dark-blue dash-dotted lines denote nonplanar gluons 
  (lying outside the tube); grey dotted lines indicate the contours
  of the cylinder.
  Diagram (a) is cylindric, has three color loops, and is of order
  $\alpha_s^3 N_c^3$; diagrams (b,c) represent cylinders with one
  gluonic handle, have only one color loop, and are of order
  $\alpha_s^3 N_c$.}
\end{figure}

We now add one more gluon. As an example, we consider adding a gluon
in the diagram of Fig.~\ref{Fig:m1}(a): The diagram of
Fig.~\ref{Fig:m2}(a) is a cylinder diagram and has the same order in
$N_c$ as the diagrams of Fig.~\ref{Fig:m1}.
The diagrams of Fig.~\ref{Fig:m2}(b,c) have the topology of a
cylinder with one handle: Adding one handle reduces the number of
color loops by one and adds one power of $\alpha_s$; the two diagrams
get a reduction factor $1/N^2_c$.  

\begin{figure}[!hb]
\includegraphics[height=3.1cm]{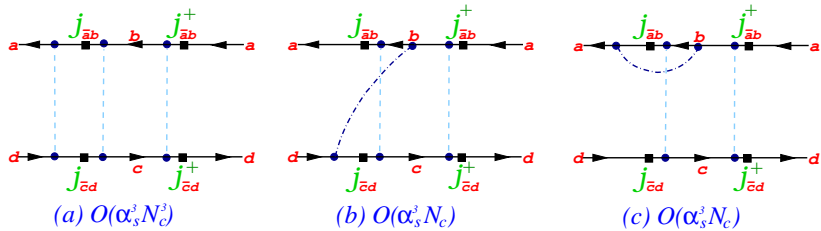}
\caption{\label{Fig:m2D}
Redrawing the cylinder diagrams of Fig.~\ref{Fig:m2} as planar
diagrams (with handles).}
\end{figure}

It is easy to establish the connection between the cylinder diagrams
of Fig.~\ref{Fig:m2} and the planar diagrams of Fig.~\ref{Fig:m2D}:
we break the quark lines $a$ and $d$ and put the two separated points
of each line at $-\infty$ and $+\infty$, respectively. Then the
classification planar/nonplanar gluon exchanges becomes obvious. When
calculating the color factors, one has to take into account that the
left and the right ends of the line $a$ ($d$) are in fact joined
together and form a color loop. 

\subsection{Recombination 4-point Green function}
\label{s2B}

\begin{figure}[h]
\includegraphics[height=4cm]{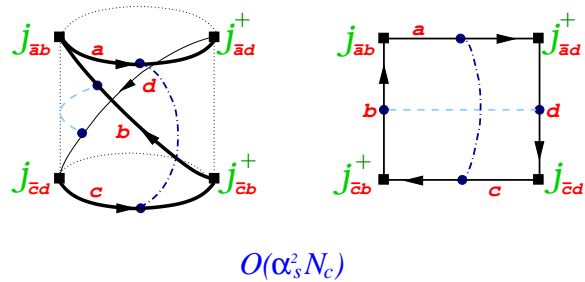}
\caption{\label{Fig:m3}
  $N_c$-leading $O(\alpha_s^2)$ diagram for the recombination amplitude
  appropriate for flavor-exotic tetraquarks.}
\end{figure}

The $N_c$-leading tetraquark-phile diagram for the recombination
Green function is shown in Fig.~\ref{Fig:m3}: it is a cylinder with
one handle, has one color loop, and is of order $\alpha_s^2 N_c$. 
Interestingly, there are no tetraquark-phile diagrams without handles
in the recombination channel. 

\section{Comparison of direct and recombination Green functions}
\label{s3}

Applying our criteria for tetraquark-phile diagrams, one finds that
the $N_c$-leading direct and recombination diagrams have 
different large-$N_c$ behaviors. If tetraquark poles emerge at all,
they should emerge in the $N_c$-leading tetraquark-phile diagrams 
(any different setup is difficult to justify from the perspective of
bound-state equations for tetraquarks).  
Then, one needs two narrow tetraquark states $T_A$ and $T_B$, each
decaying via one preferred meson-meson channel, in order to satisfy
the consistency conditions between {\cal D} and {\cal R} Green
functions \cite{lms_prd,lms_epjc}: 
\begin{eqnarray}
\label{T4b2}
&& A(T_A\to M_{\bar ab}M_{\bar cd})=O(N_c^{-1}),\qquad A(T_A\to
M_{\bar ad}M_{\bar cb})=O(N_c^{-2}),\nonumber\\
&& A(T_B\to M_{\bar ab}M_{\bar cd})=O(N_c^{-2}),\qquad A(T_B\to
M_{\bar ad}M_{\bar cb})=O(N_c^{-1}).
\end{eqnarray}
If the bound states exist, their widths $\Gamma(T_{A,B})$ will be
determined by the dominant channels, which yields
$\Gamma(T_{A,B})=O(N_c^{-2})$, thus confirming the narrow-width
property of the tetraquark candidate states.

\begin{figure}[ht]
\begin{center}
\includegraphics[height=4.0cm]{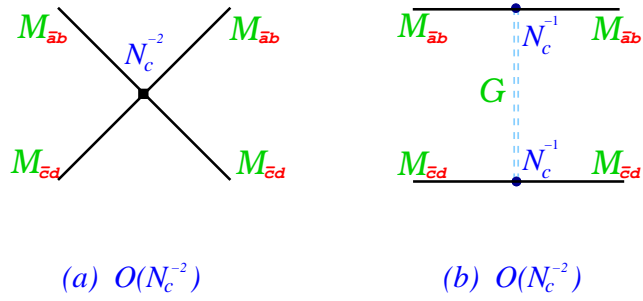}
\caption{Effective tree-level meson-meson interactions in the
direct channels: (a) Four-meson contact term. (b) Glueball
exchange.}
\label{Fig5} 
\end{center}
\end{figure}

The $N_c$-matching conditions also allow us to deduce the
properties of the effective tree-level meson-meson interactions.
The direct-channel $N_c$-leading connected diagrams are
OZI-suppressed \cite{largeNc2,coleman} and the corresponding
effective meson-meson interactions come out to be of order $1/N_c^2$,
resulting either from contact terms (Fig. \ref{Fig5}(a)) or from
glueball exchanges (Fig. \ref{Fig5}(b)) \cite{lms_epjc}.
\par
\begin{figure}[ht]
\begin{center}
\includegraphics[height=4.0cm]{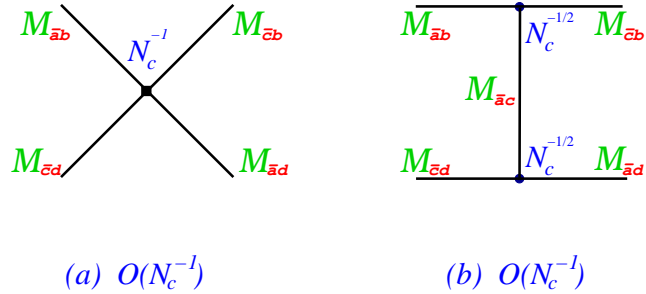}
\caption{Effective tree-level meson-meson interactions in the
recombination channel: (a) Four-meson contact term. (b) Meson
exchange (to be completed with a similar diagram in the
$u$-channel).}
\label{Fig6} 
\end{center}
\end{figure}

On the other hand, the recombination-channel interactions are of
the generic order $1/N_c$, resulting either from contact terms
(Fig. \ref{Fig6}(a)) or from meson exchanges (Fig. \ref{Fig6}(b))
\cite{lms_epjc}. (They are provided by the $N_c$-leading diagrams
of the recombination channel.)
\par
Taking into account the above properties and Eqs. (\ref{T4b2}),
one deduces the dominant structure of each of the two
tetraquarks: $T_A$ has the structure $(\bar ad)(\bar cb)$,
while $T_B$ has the structure $(\bar ab)(\bar cd)$, both of them
being the product of two color-singlet clusters; their dominant
decay channel proceeds through the recombination (quark-exchange)
process, rather than through the dissociation one.

However, in the diquark-antidiquark mechanism of the tetraquark
formation \cite{Schafer:1993ra,Jaffe:2003sg,Nussinov:2003ex,
Shuryak:2003zi,Maiani:2004vq}, one disposes of one flavor-exotic
combination $(\bar a\bar c)(bd)$, in the form of the product of
color-antisymmetric representations, which leads to a contradiction
with the above requirement of the existence of two tetraquarks. 
One thus is led to conclude that in large-$N_c$ QCD flavor-exotic
compact narrow tetraquarks might not exist. 

This conclusion rests on the observed different behaviors of
the tetraquark-phile contributions to direct and recombination
Green functions. Can one formulate different consistent criteria
for selecting tetraquark-phile diagrams? It is conceivable that for
some dynamical reasons, tetraquarks do not necessarily contribute
to the generic leading diagrams that have been taken into account.
The authors of \cite{maiani2} consider such a possibility by
imposing more stringent selection rules. These are based on two
main assumptions: (i) Tetraquark-phile diagrams have a nonplanar
topology with one gluonic handle. (ii) Only one class of diagrams,
either {\cal D} or {\cal R}, contributes to the tetraquark formation.
For phenomenological reasons, it is channel {\cal D} that is chosen
as admissible for tetraquark emergence. Then a single tetraquark
may accommodate the consistency conditions, with a coupling of
order $N_c^{-2}$ to the two sets of available meson pairs. 

Without intending to discard the possibility of a selection mechanism
as described in \cite{maiani2}, which demands, however, a more
detailed investigation on dynamical grounds, we would like to
draw attention to one argument that does not seem well founded.

\begin{figure}[ht]
\begin{center}
\includegraphics[height=4.0cm]{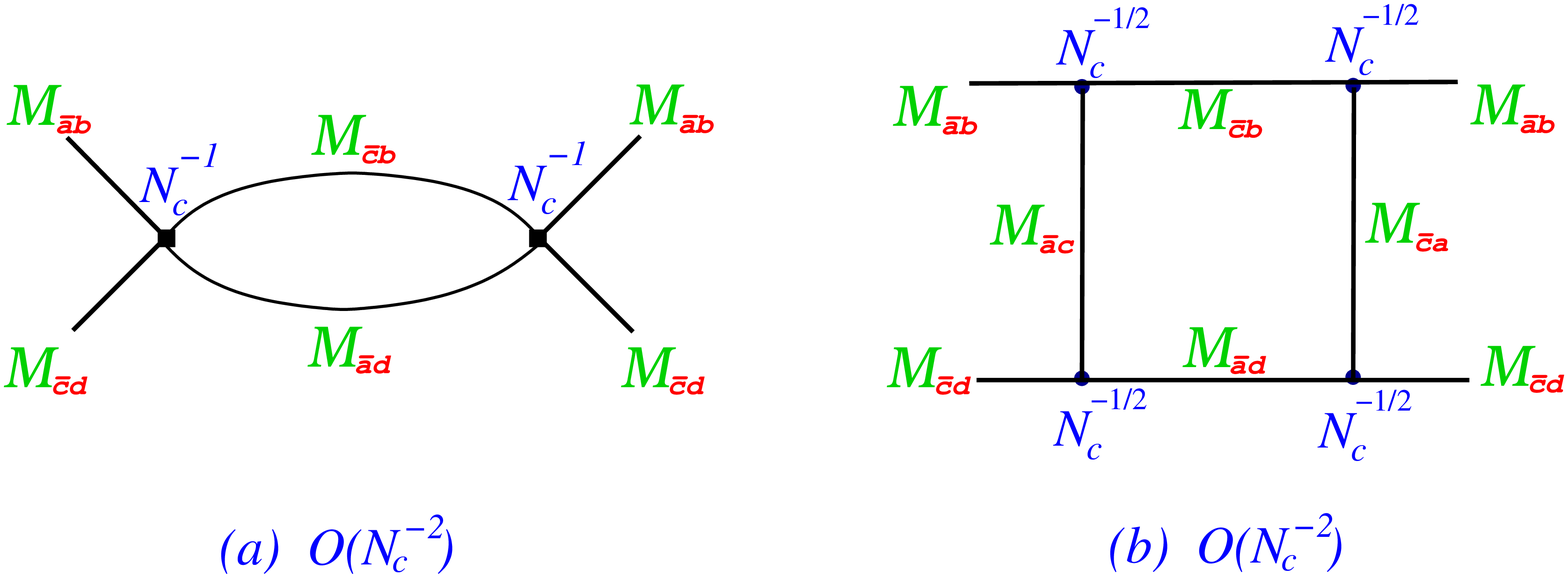}
\caption{Contributions of two-meson intermediate states to
the meson-meson scattering amplitude in a direct channel.
The intermediate states are those produced by the recombination
process at lower order.}
\label{Fig7} 
\end{center}
\end{figure}

The main justification in \cite{maiani2} of imposing on the
channel {\cal D} tetraquark-phile diagrams to have a handle is
based on the assertion that {\cal D}-type planar diagrams do
not describe mutual interactions of meson or $(\bar qq)$ pairs.
However, this is contradicted by the existence of two-meson
intermediate states contributing through planar diagrams to
meson-meson scattering. First, the $N_c$-leading
diagrams of the {\cal R} channel, which do not have four-quark
singularities, still contribute to the effective meson-meson
interaction through the diagrams of Fig. \ref{Fig6}, the global
coupling being of order $N_c^{-1}$. Second, the unitarity
condition requires that the diagrams of Fig. \ref{Fig6} generate
meson loop diagrams of the type of Fig.~\ref{Fig7}, which are
genuine parts of the meson-meson scattering amplitude in channel
{\cal D}. They are of order $N_c^{-2}$, i.e., of the same order
as the leading planar diagrams of channel {\cal D}. This could
not happen if their underlying QCD diagrams were not of the
planar type.
\par
\begin{figure}[ht]
\begin{center}
\includegraphics[height=5.0cm]{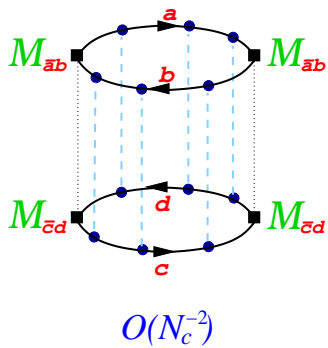}
\caption{A typical cylindric QCD diagram (cf. Fig.~\ref{Fig:m2}a)
contributing to the two-meson intermediate states in the meson-meson
scattering amplitude in a direct channel. Gluon exchanges around the
corners of the loops, going from one line of a loop to the other
line, are parts of the external meson states and are not drawn.}
\label{Fig8} 
\end{center}
\end{figure}

A typical such diagram is presented in Fig. \ref{Fig8}. The
intermediate states, obtained from a vertical cut are precisely
those corresponding to the quark-exchange process. Here, color
rearrangement plays a physical role by converting the pair of
initial mesons into the other pair.
\par
\par
This shows that the leading-order planar diagrams of channel
{\cal D} are not physically empty and describe a part of the
meson-meson scattering process. The question whether a compact
tetraquark pole may emerge from such a process or not still remains
a relevant issue.
Meson-meson interactions are expected to be short-range and if
a tetraquark pole exists in the corresponding scattering
amplitude, as a bound state or a resonance, it should be loosely
bound and would presumably correspond to a molecular-type state.
This possibility is examined in Sec. \ref{s4}.

The possible existence of a hidden dynamical mechanism which
favors the emergence of compact tetraquarks in fully
exotic channels is an open question and deserves further study.
We also emphasize that the conclusions obtained in the main
part of this section apply only to the fully exotic case
(four different quark flavors). For systems with a smaller
number of quark flavors, additional QCD diagrams, not present
in the fully exotic case, may invalidate some of the results
obtained above and may allow for the existence of one tetraquark 
\cite{lms_prd,lms_epjc}.

\section{Molecular states at large $N_c$}
\label{s4}

Tetraquarks may also have a molecular structure, resulting from
meson-meson interactions and existing either in the form of
bound states or of resonances \cite{Voloshin:1976ap,Bander:1975fb,
DeRujula:1976zlg,Tornqvist:1993ng,Amsler:2004ps,Swanson:2006st,
Guo:2017jvc}. Meson-meson interactions are generally formulated in
the form of effective Lagrangians or of empirical potentials. In
the large-$N_c$ limit, these interactions are expected to scale as
$1/N_c$ \cite{largeNc2,coleman}.
 
In the case of mesons made of light quarks ($u,d,s$) and, in
particular, involving the lightest pseudoscalar mesons, chiral
perturbation theory (ChPT) \cite{Gasser:1983yg,Gasser:1984gg}
provides the general effective Lagrangian suited for the description
of the corresponding interactions at low energies. An extension of
the energy domain of validity of ChPT is done with the aid of the
unitarization condition of the scattering amplitudes, together with
the use of dispersion relations \cite{Pelaez:2006nj,Pelaez:2015qba}.

In sectors involving heavy quarks, the masses of the latter
introduce new scales in the system, which must be taken into account.
The heavy mesons, in addition to their contact-type interactions,
also interact through light-meson exchanges of Yukawa type 
\cite{Amsler:2004ps,Swanson:2006st,Valderrama:2012jv,
Karliner:2016ith} (cf. Fig.~\ref{Fig6}(b), where the heavy quarks
correspond to those denoted $b$ and $d$). The latter interactions
provide additional opportunities for the emergence of bound states
or of resonances.
A more systematic study can be done with the use of Heavy Quark
Effective Field Theory and the associated spin and flavor symmetries
\cite{Guo:2017jvc,Georgi:1990um,Wise:1992hn,Neubert:1993mb,
Luke:1996hj,AlFiky:2005jd,Baru:2018qkb}.

\begin{figure}[ht]
\begin{center}
\includegraphics[height=2.2cm]{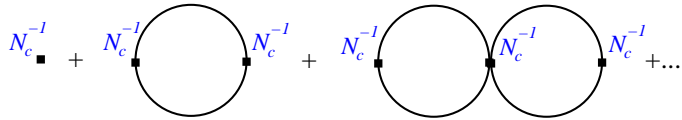}
\caption{Chain of bubble diagrams participating in the
realization of elastic unitarity and in the creation of
a bound state or of a resonance pole; the $N_c$ dependence of the
effective meson-meson interaction has been factored out.}
\label{Fig9} 
\end{center}
\end{figure}

The possible dynamical emergence of bound states or of resonances
from meson-meson interactions is studied by summing in the
evaluation of the scattering amplitude chains of diagrams of similar
structure; Fig. \ref{Fig9}, where we have explicitly factored out
the $N_c$ dependence of the effective coupling constant, schematically
displays the summation of bubble diagrams.
These diagrams are generally divergent and, accordingly, the
effective coupling constants undergo renormalization
\cite{Luke:1996hj,Weinberg:1991um,Jackiw:1991}. According to the
signs of the renormalized coupling constants, which might be
determined from other experimental data, a bound state or a
resonance pole may emerge.

The important qualitative feature of the resultant dynamical pole
is that its mass squared, in the case of a resonance, or binding
energy, in the case of a bound state, are essentially proportional,
up to small corrections, to the inverse of the effective coupling
constant \cite{Luke:1996hj,Weinberg:1991um,Jackiw:1991}. Therefore,
assuming that the renormalized effective coupling constant remains,
as the bare one, inversely proportional to $N_c$, the resonance
mass will be pushed towards infinity with a broad width, while the
bound state will disappear from the spectrum.

In the case of light quarks, a detailed study of the problem
has been presented in \cite{Pelaez:2006nj,Pelaez:2015qba}. The
general result is that, in the scalar-isoscalar $s$-channel of
$\pi\pi$ scattering, a dynamical resonance, corresponding to the
observed $f_0(500)$ resonance, emerges, having a dominant structure
of two quarks and two antiquarks, in distinction from the
ordinary mesons \cite{Jaffe:2008zz}. This result has also been
confirmed by a direct solution of the four-quark Bethe-Salpeter
equation \cite{Eichmann:2015cra}. At large $N_c$,
the mass and the width of the resonance behave as $\sqrt{N_c}$,
as expected from the general qualitative features outlined above.
Generalization of the calculations with other combinations of the
light quarks is expected to provide similar qualitative
conclusions.

In the case of sectors involving heavy quarks, the above
conclusions would remain true in the formal limit of $N_c$ going
to infinity, but for finite values of $N_c$ the observable
effects might be less striking, since the mass gap between the
two-meson threshold and the resonance position would be relatively
reduced as compared to the light-quark-sector case.

Can molecular-type tetraquarks contribute to the large-$N_c$
analysis of Green functions? The answer is negative, since
the emergence of a molecular-type pole necessitates the summation
of a chain of diagrams with different orders in $1/N_c$
(Fig. \ref{Fig9}). This is in contrast to the case of compact
tetraquarks, where the summation of planar diagrams is done at
the same order in $N_c$ together with the creation of the pole;
this makes possible the matching of the pole contributions in the
Green function, on the one hand, and in the Feynman diagrams, on
the other. For molecular-type tetraquarks, the latter contribute
to Feynman diagrams at leading order in $N_c$ only through their
contact terms and one-meson exchange terms (Fig. \ref{Fig6}) and
therefore a matching-type analysis is not possible.
In that case, one has to deduce the properties of the tetraquark
from the explicit summation itself.

A particular attention should be paid to the case of molecular-type
bound states, also called deuteron-like, which may emerge very close
to the two-meson threshold, with an unnaturally small binding energy  
\cite{Guo:2017jvc,Weinberg:1965zz}. They are also characterized by a
large (negative) value of the S-wave scattering length, much greater
than the natural scale provided by the physical parameters of the
system, and exhibiting universality properties \cite{Braaten:2004rn}.
According to the general properties of the emergence of dynamical
poles in the scattering amplitude as outlined above, these bound
states might appear only in the strong-coupling limit of the
effective theory, while the large-$N_c$ limit drives the theory to
its weak-coupling limit. Therefore, in the formal limit of large
$N_c$, one also predicts the diappearance of these states. In general,
the details of the creation mechanism of these states not being
well known, it is admitted that underlying fine-tuning processes  
might be at work for their existence \cite{Braaten:2004rn}. This
would mean that they are very sensitive to variations of their
physical parameters and, in particular, of $N_c$. Results obtained
at large-$N_c$ might not correctly describe their physical properties
at finite values of $N_c$. 

Coming now back to the case of flavor-exotic tetraquarks, we
recall that the direct-channel effective meson-meson interactions
are actually of order $1/N_c^2$ (cf. Fig. \ref{Fig5}), i.e., they
are much weaker than in the generic case ($\sim 1/N_c$). The
recombination-channel interactions remain of order
$1/N_c$ (cf. Fig. \ref{Fig6}); however, since they represent
off-diagonal-type contributions in a coupled-channel
formalism, their effective contributions to the resonance
or the bound-state pole formation will still be as in the
direct-channel case. Therefore, the possibly existing
resonance-pole positions will be pushed even more strongly to
infinity than in the generic cases, while bound-state poles will
be absent from the spectrum.

In conclusion, narrow-width molecular-type tetraquarks, with masses
that remain fixed at large $N_c$, are not generally expected to occur
in flavor-exotic sectors. Assuming that the continuation to finite
values of $N_c$ remains a smooth operation in the theory, this
statement would still be valid in the physical world, except possibly
in the particular case of a bound state lying very close to the
two-meson threshold.

\section{Conclusions}
\label{s5}

We have considered, in the large-$N_c$ limit of QCD, the possibility
of the existence of narrow four-quark states of an exotic flavor
content, involving four quarks of different flavors (that requires
two quarks and two antiquarks as a minimal parton configuration).
The two cases of compact and molecular tetraquarks have been examined.

Compact tetraquarks are the genuine candidates for the quest for
narrow-width states at large $N_c$. In the sectors of flavor-exotic
states, the consistency constraints, coming from the direct 
and recombination (or quark-exchange) type channels,
require the existence of two different tetraquarks, each having
a structure made of two color-singlet clusters or mesons, and
decaying in a preferred two-meson channel, fixed by the dominance
of the recombination-type effective interaction. On the other hand,
the formation mechanism of tetraquarks through a primary formation
of diquarks and antidiquarks predicts the existence of one
tetraquark, decaying with equal weights, up to small corrections,
into the two different two-meson channels. This contradiction
suggests that compact tetraquarks do not exist in flavor-exotic
sectors, unless some hidden dynamical mechanism favors their
emergence \cite{Brodsky:2014xia,Maiani:2017kyi,maiani2}.

Molecular tetraquarks, because of the weakening of the effective
meson-meson interactions at large $N_c$, might only exist as
resonances with masses and widths that increase like $\sqrt{N_c}$.
In the case of the presence of heavy mesons, the mass gap between
the resonance position and the two-meson thresholds might be
substantially reduced at finite values of $N_c$.
In the flavor-exotic case, the effective interactions are
much weaker than in the generic cases, and, because of this feature,
the masses of the possibly existing resonances are repelled to
higher values. Therefore, at large $N_c$, no molecular-type
tetraquarks, with fixed masses and narrow widths, are expected to
emerge. An exceptional case might occur, at finite $N_c$, with the
emergence of a single bound state lying very close to the two-meson
threshold.

Up to now, experimental data, as well as lattice calculations, do
not provide evidence for the existence of flavor-exotic tetraquarks
in sectors involving one heavy quark, $c$ or $b$. Flavor-exotic
sectors involving two heavy quarks, $c$ and $b$, seem to be yet
unexplored. Therefore, experimental data and lattice calculations
for these sectors would be of great help for the understanding
of the underlying dynamics of QCD.

\vspace{.5cm}
\noindent{\it{\bf Acknowledgements.}} 
The authors thank V.~Anisovich, T.~Cohen, L.~Gladilin, M.~Knecht,
B.~Moussallam, and B.~Stech for valuable discussions.
D.~M.~acknowledges support from the Austrian Science Fund (FWF),
project~P29028.

\end{document}